\documentclass[12pt]{article}
\usepackage{amsmath,amssymb,bm,epsfig}

\renewcommand{\thefootnote}{\fnsymbol{footnote}}

\begin{document}

\title{
\begin{flushright}
\begin{minipage}{0.2\linewidth}
\normalsize
WU-HEP-12-05 \\*[50pt]
\end{minipage}
\end{flushright}
{\Large \bf 
The Higgs boson mass in a natural MSSM with 
nonuniversal gaugino masses at the GUT scale
\\*[20pt]}}

\author{Hiroyuki~Abe\footnote{
E-mail address: abe@waseda.jp}, \ 
Junichiro~Kawamura\footnote{
E-mail address: junichiro-k@ruri.waseda.jp} \ and \ 
Hajime~Otsuka\footnote{
E-mail address: hajime.13.gologo@akane.waseda.jp
}\\*[20pt]
{\it \normalsize 
Department of Physics, Waseda University, 
Tokyo 169-8555, Japan} \\*[50pt]}

\date{
\centerline{\small \bf Abstract}
\begin{minipage}{0.9\linewidth}
\medskip 
\medskip 
\small
We identify a parameter region where the mass of the lightest 
CP-even Higgs boson resides in $124.4-126.8$ GeV, and 
at the same time 
the degree of tuning a Higgsino-mass parameter (so-called 
$\mu$-parameter) is relaxed above $10$ \% 
in the minimal supersymmetric standard model (MSSM) 
with soft supersymmetry breaking terms, 
by solving the full set of one-loop renormalization group 
equations numerically. It is found that certain nonuniversal 
values of gaugino-mass parameters at the so-called grand 
unification theory (GUT) scale $\sim 10^{16}$ GeV are 
important ingredients for the MSSM to predict, without a 
severe fine-tuning, the Higgs boson mass $\sim 125$ GeV 
indicated by recent observations at the Large Hadron Collider. 
We also show a typical superparticle spectrum 
in this parameter region. 
\end{minipage}
}

\begin{titlepage}
\maketitle
\thispagestyle{empty}
\clearpage
\tableofcontents
\thispagestyle{empty}
\end{titlepage}

\renewcommand{\thefootnote}{\arabic{footnote}}
\setcounter{footnote}{0}

\section{Introduction}

The low-energy supersymmetry is one of the most promising 
candidates for a new physics beyond the standard model (SM) 
of elementary particles due to the absence of quadratic 
divergences. The supersymmetric partners of SM particles 
cancel the radiative corrections to the mass of Higgs bosons, 
then protect the electroweak (EW) scale $M_{\rm EW} \sim 10^2$ GeV 
against the huge corrections. 
The lightest supersymmetric particle (LSP) can be a candidate 
for the dark matter required from cosmological observations. 
Moreover, the minimal supersymmetric standard model (MSSM) 
predicts that three gauge coupling constants are unified at 
a high-energy scale around 
$M_{\rm GUT} \simeq 2 \times 10^{16}$ GeV, 
and the electroweak symmetry is broken in a wide range 
of the parameter space of the MSSM with soft supersymmetry 
breaking terms, due to logarithmic radiative corrections 
that cause sizable running of parameters from 
$M_{\rm GUT}$ to $M_{\rm EW}$. 
(See, for a review, Ref.~\cite{Martin:1997ns}.) 

On a stable vacuum where the EW symmetry is broken 
successfully, the mass of $Z$-boson is determined by 
\begin{eqnarray}
m_Z^2 &=& 
\frac{\left| m_{H_d}^2(M_{\rm EW})
-m_{H_u}^2(M_{\rm EW}) \right|}{
\sqrt{1-\sin^2 (2 \beta)}} 
-m_{H_u}^2(M_{\rm EW})-m_{H_d}^2(M_{\rm EW}) 
-2 \left| \mu(M_{\rm EW}) \right|^2 
\nonumber \\
&\simeq& -2 \left| \mu(M_{\rm EW}) \right|^2 
-2 m_{H_u}^2(M_{\rm EW}), 
\label{eq:mz}
\end{eqnarray}
where $\mu(M_{\rm EW})$ and $m_{H_u}(M_{\rm EW})$ 
are supersymmetric and soft supersymmetry breaking masses, 
respectively, for the Higgs field evaluated at $M_{\rm EW}$. 
The Higgsino mass is also given by $\mu$ and we refer 
to this parameter as $\mu$-parameter. 
The above mentioned radiative EW breaking roughly means 
$m_{H_u}^2(M_{EW})<-|\mu(M_{EW})|^2<0$ 
even with 
$m_{H_u}^2(M_{{\rm GUT}})>0$ 
at $M_{\rm GUT}$. 
The observed value $m_Z=91.2$ GeV indicates 
$|\mu(M_{\rm EW})| \sim |m_{H_u}(M_{\rm EW})| \sim M_{\rm EW}$, 
otherwise a fine-tuning is required between parameters 
$\mu$ and $m_{H_u}$. 
One of the guiding principles toward a more fundamental 
theory beyond the MSSM can be provided by an argument of 
the naturalness. Because the mass parameters $\mu$ and 
$m_{H_u}$ should have essentially different origins 
from the viewpoint of supersymmetry, there is no reason 
that these parameters are closely related. 

However, from the recent results in the search for Higgs and 
supersymmetric particles at the Large Hadron Collider (LHC), 
the mass of scalar quarks (squarks) in the first and the 
second generation is indicated above about $1.3$ TeV~\cite{:2012rz}, 
and the allowed region of the mass of the lightest CP-even 
Higgs boson has been reported in between $124.4$ and 
$126.8$ GeV~\cite{:2012gk}. 
The latter implies a large radiative correction for the 
mass of the Higgs boson which is lighter than the 
$Z$-boson at the tree-level~\cite{Okada:1990vk}. 
In this case the mass of 
scalar top quarks must be much heavier than $M_{\rm EW}$ 
which dominantly contribute to such a correction due 
to a large top Yukawa coupling. These observations 
indicate that the mass scale of soft supersymmetry 
breaking parameters tends to be much larger than $m_Z$, 
which in general cause the fine-tuning problem 
mentioned above. 

It was pointed out in Ref.~\cite{Abe:2007kf} 
that certain nonuniversal gaugino masses at $M_{\rm GUT}$ 
relax the  degree of the above mentioned fine-tuning in the MSSM. 
In this paper, we update the analysis based on the 
recent experimental data.\footnote{Similar analysis 
is performed recently in Ref.~\cite{Antusch:2012gv}.} 
Because the latest Higgs mass bound shown above indicates 
a larger value of $\tan \beta$ than the value adopted 
in Ref.~\cite{Abe:2007kf}, 
in the following analysis, we solve the full set of 
one-loop renormalization group equations (RGEs), 
in contrast to the previous analysis in Ref.~\cite{Abe:2007kf} 
where all the Yukawa (and scalar trilinear) couplings are 
neglected except for those involving only (scalar) top quarks. 

The following sections are organized as follows. 
In Sec.~\ref{sec:mssm}, we roughly estimate the effects 
of certain nonuniversal gaugino masses on the mass of 
the lightest CP-even Higgs boson and on the degree 
of tuning the $\mu$-parameter. 
Based on these implications, in Sec.~\ref{sec:numerical}, 
we perform a full numerical analysis at the one-loop 
level and identify a parameter region where the Higgs 
mass resides in $124.4-126.8$ GeV and the degree of 
tuning is relaxed above 10 \% at the same time. Finally, 
Sec.~\ref{sec:conclusion} is devoted to 
conclusions and discussions. 
In Appendix~\ref{app:bc}, we show the boundary conditions 
for RGEs at the GUT scale adopted in this paper. 
The relevant RGEs for analyzing the Higgs mass are 
exhibited in Appendix~\ref{app:relRGEs}.

\section{Implications of nonuniversal gaugino masses}
\label{sec:mssm}

The superpotential and soft supersymmetry breaking terms 
in the MSSM are shown in Eqs.~(\ref{eq:wmssm}) and 
(\ref{eq:softterms}) in Appendix~\ref{app:bc}, respectively. 
We use the notations and the conventions adopted in 
Appendix~\ref{app:bc} throughout this paper. 

The radiative corrections to the Higgs mass are 
dominated by loops of scalar top quarks. 
With an approximation that the mass eigenstates 
of top squarks are nearly degenerate, the mass of the 
lightest CP-even Higgs boson is evaluated as~\cite{Carena:1995wu} 
\begin{eqnarray}
m_h^2 &\simeq& 
m_Z^2 \cos^2 (2 \beta) \left( 
1-\frac{3}{8 \pi^2} \frac{\overline{m_t}^2}{v^2} t \right) 
\nonumber \\ &&
+\frac{3}{4 \pi^2} \frac{\overline{m_t}^4}{v^2} \left[ 
\frac{1}{2}X_t + t + \frac{1}{16 \pi^2} 
\left( \frac{3}{2} \frac{\overline{m_t}^2}{v^2} 
-32 \pi \alpha_3 \right) 
\left( X_t t + t^2 \right) \right], 
\label{eq:mh}
\end{eqnarray}
with 
\begin{eqnarray}
X_t &=& \frac{2 \tilde{A}_t^2}{M_{\rm st}^2} 
\left( 1-\frac{\tilde{A}_t^2}{12 M_{\rm st}^2} \right), 
\qquad 
\left\{ \begin{array}{rcl}
\tilde{A}_t &\equiv& A_t(m_Z)-\mu(m_Z) \cot \beta \\[5pt]
M_{\rm st}^2 &\equiv& \sqrt{m_{U_3}^2(m_Z)\,m_{Q_3}^2(m_Z)} 
\end{array} \right., 
\nonumber
\end{eqnarray}
where 
$\overline{m_t} = 165$ GeV 
is the top quark mass, 
$y_t \equiv y^u_{33}$ is the top Yukawa coupling, 
$a^u_{33} \equiv y_t A^u_{33} \equiv y_t A_t $ 
is the scalar trilinear coupling 
involving only top squarks, 
$m_{Q_3}^2 \equiv (m_Q^2)_{33}$ 
($m_{U_3}^2 \equiv (m_U^2)_{33}$) is the left-(right-)handed 
top squark mass square, and $t=\ln (M_{\rm st}^2/\overline{m_t}^2)$. 
Here $v = 174$ GeV is related to the VEVs of up- and 
down-type Higgs fields $h_u$ and $h_d$ as 
$v_u = v \sin \beta$ and $v_d = v \cos \beta$, respectively. 
As is well known, the value of $X_t$ is maximized with 
$|\tilde{A}_t| \sim \sqrt{6} M_{\rm st}$, and then 
the loop corrections are enhanced in Eq.~(\ref{eq:mh}). 

If all the Yukawa (and scalar trilinear) couplings are 
neglected except for those involving only (scalar) top quarks, 
the one-loop RGEs show that the soft supersymmetry breaking 
parameters evaluated at the $Z$-boson mass scale $m_Z$ 
are related to those at 
$M_{\rm GUT}$ as 
\begin{eqnarray}
m_{Q_3}^2(m_Z) &\simeq& 
-0.02 M_1^2+0.38 M_2^2-0.02 M_1 M_3
-0.07 M_2 M_3+5.63 M_3^2 
\nonumber \\ &&
+(0.02 M_2+0.09 M_3-0.02 A_t)A_t
\nonumber \\ && 
-0.14 m_{H_u}^2+0.86 m_{Q_3}^2-0.14 m_{U_3}^2, 
\nonumber \\
m_{U_3}^2(m_Z) &\simeq& 
0.07 M_1^2-0.01 M_1 M_2-0.21 M_2^2
-0.03 M_1 M_3-0.14 M_2 M_3+4.61 M_3^2 
\nonumber \\ &&
+(0.01M_1+0.04 M_2+0.18 M_3-0.05 A_t) A_t
\nonumber \\ && 
-0.27 m_{H_u}^2-0.27 m_{Q_3}^2+0.73 m_{U_3}^2, 
\nonumber \\
A_t(m_Z) &\simeq& 
-0.04 M_1-0.21 M_2-1.90 M_3+0.18 A_t, 
\label{eq:spsatmz}
\end{eqnarray}
where 
the soft parameters without any arguments 
in the right-handed sides represent those 
evaluated at $M_{\rm GUT}$. 
Here the numerical values of the MSSM gauge couplings 
and the top Yukawa coupling are chosen in such a way that 
these values become the observed ones at low energies, 
and we take $\tan \beta =15$ as declared at the end 
of Appendix~\ref{app:bc}. 
The numerical values of the coefficients 
in Eq.~(\ref{eq:spsatmz}) suggest 
$|\tilde{A}_t|/M_{\rm st}<1$ for soft parameters 
of the same orders of magnitude at $M_{\rm GUT}$, 
without any cancellation between terms in the 
right-handed sides of Eq.~(\ref{eq:spsatmz}). 

This insists that a certain cancellation is 
required in Eq.~(\ref{eq:spsatmz}) in order to 
enhance radiative corrections in Eq.~(\ref{eq:mh}) 
for obtaining a heavier Higgs mass. 
We find that the gluino mass-square term with the 
largest coefficient can be cancelled by the wino 
mass-square term in $m_U^2(m_Z)$ with the ratio 
$M_2/M_3 \sim \sqrt{4.6/0.21} \sim 4.8$, 
and then $0<m_U(m_Z)<|A_t(m_Z)|$ is realized. 
This phenomenon can be understood as follows. 
The gaugino masses and soft scalar masses act as 
positive and negative driving forces, respectively, 
in the renormalization group evolution of soft scalar 
mass square from the GUT to the EW scale. 
The relevant RGEs are shown in Appendix~\ref{app:relRGEs}. 
Therefore, 
the mass square of the left-handed scalar quark 
$m_{Q}^2$ tends to increase for a larger wino mass 
$M_2/M_3 >1$, and then the mass square of the 
right-handed squark $m_{U}^2$ tends to decrease 
with the increasing $m_{Q}^2$. 
The averaged top squark mass $M_{\rm st}$ decreases 
because the increasing contribution from $m_{Q}^2$ 
is dominated by the decreasing one from $m_{U}^2$. 
As we can estimate in Eq.~(\ref{eq:spsatmz}), 
the contribution from the wino mass satisfying 
$2 \lesssim M_2/M_3 \lesssim 5$ to reduce $m_U^2$ 
dominates the one to increase $m_Q^2$, 
and then $|\tilde{A}_t|/M_{\rm st}>1$ is achieved. 

\begin{figure}[t]
\hfill 
\includegraphics[width=0.45\linewidth]{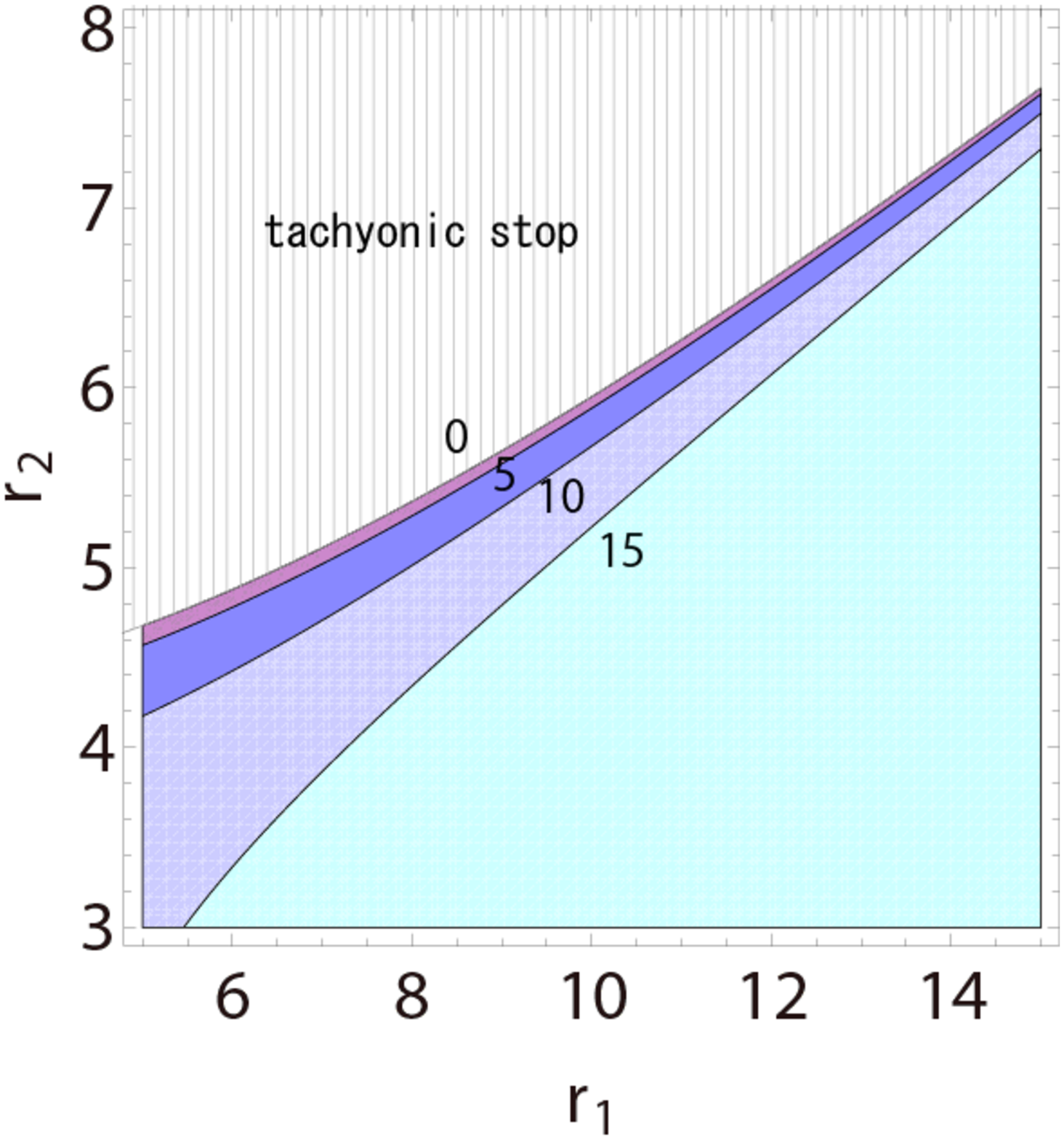}
\hfill 
\includegraphics[width=0.45\linewidth]{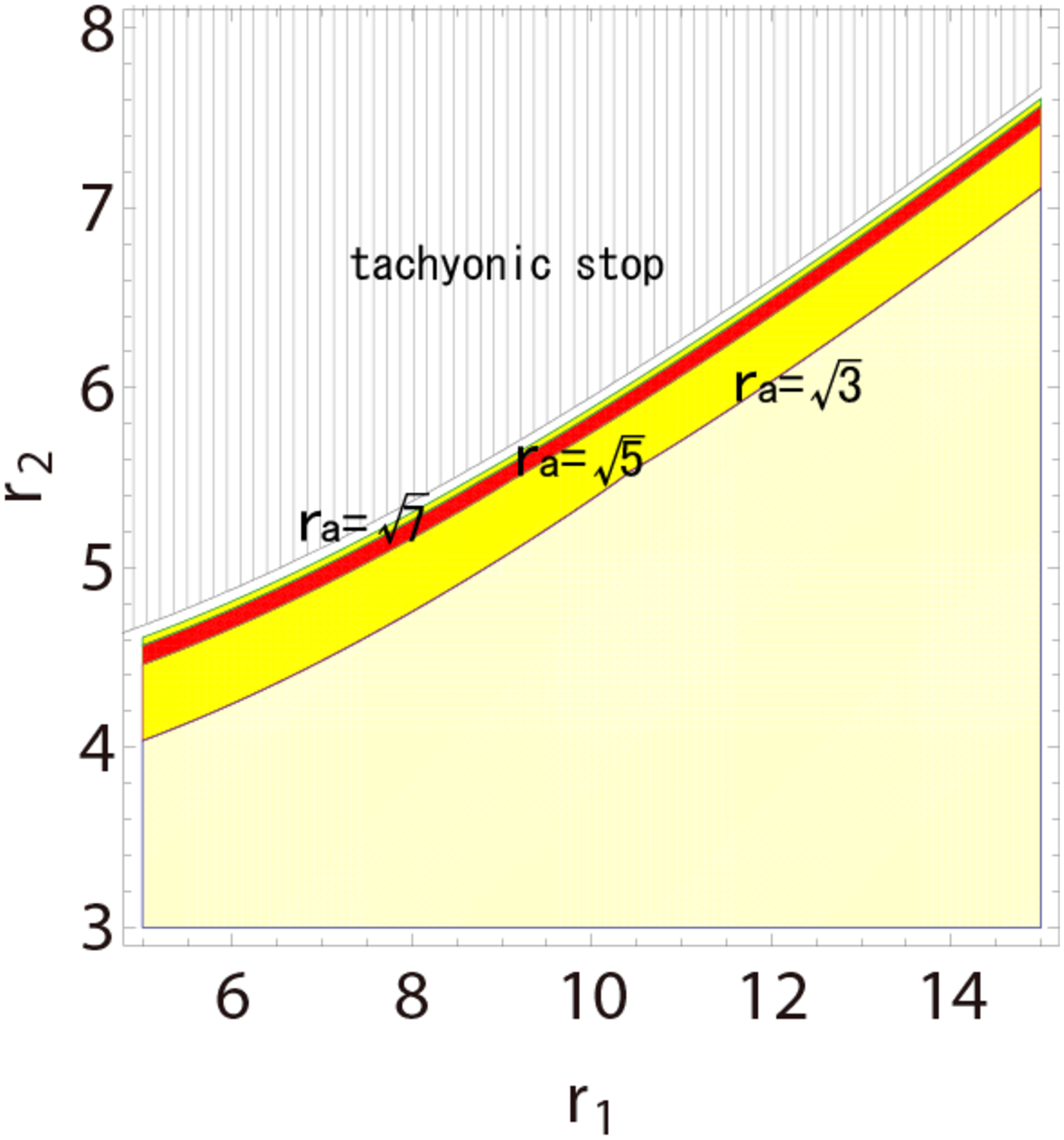}
\hfill 
\caption{Contours of 
$M_{\rm st}/m_{Q_3}$ (the left panel) and those of 
$r_a \equiv |\tilde{A}_t|/M_{\rm st}$ (the right panel) 
in the $(r_1,r_2)$ plane, evaluated by a numerical solution 
for the full set of one-loop RGEs with the boundary conditions 
at the GUT scale shown in Appendix~\ref{app:bc}.}
\label{fig:atmst}
\end{figure}

The above rough estimation can be verified numerically. 
Fig.~\ref{fig:atmst} shows contours of 
$M_{\rm st}/m_{Q_3}$ (the left panel) and those of 
$r_a \equiv |\tilde{A}_t|/M_{\rm st}$ (the right panel) 
in the $(r_1,r_2)$ plane, evaluated by a numerical solution 
for the full set of one-loop RGEs with the boundary conditions 
at the GUT scale shown in Appendix~\ref{app:bc}, 
where the bino-gluino and the wino-gluino mass ratios 
at the GUT scale are denoted respectively by 
\begin{eqnarray}
r_1 &\equiv& M_1/M_3, \qquad 
r_2 \ \equiv \ M_2/M_3. 
\nonumber
\end{eqnarray}
The ratios $r_1$ and $r_2$ are varied in the following analysis. 
From Fig.~\ref{fig:atmst} we find $M_{\rm st}$ decreases 
and then $|\tilde{A}_t|/M_{\rm st}$ increases 
as the wino mass parameter $M_2$ increases, 
assuring the above rough estimation.  

On the other hand, concerning about the fine-tuning, 
a smaller absolute value of the soft scalar mass square 
for the up-type Higgs $m_{H_u}^2$ is favored. 
A similar relation to the ones in Eq.~(\ref{eq:spsatmz}) 
is obtained for $m_{H_u}^2$ as 
\begin{eqnarray}
m_{H_u}^2(m_Z) &\simeq&  
- 0.01 M_1 M_2 + 0.17 M_2^2 - 0.05 M_1 M_3 
- 0.20 M_2 M_3 - 3.09 M_3^2 
\nonumber \\ && 
+ (0.02 M_1 + 0.06 M_2 + 0.27 M_3 - 0.07 A_t) A_t 
\nonumber \\ && 
+ 0.59 m_{H_u}^2 - 0.41 m_{Q_3}^2 - 0.41 m_{U_3}^2. 
\label{eq:mhuatmz}
\end{eqnarray}
The numerical values of the coefficients suggest that 
$m_{H_u}^2(m_Z)$ is of the order of soft parameters for 
those of the same orders of magnitude at $M_{\rm GUT}$, 
without any cancellation between terms in the 
right-handed side of Eq.~(\ref{eq:mhuatmz}). 

This insists again that a certain cancellation is 
required in Eq.~(\ref{eq:mhuatmz}) in order to 
realize $-m_{H_u}^2(m_Z) \sim m_Z^2$ to reduce the 
fine-tuning required by Eq.~(\ref{eq:mz}). 
We find that the gluino mass-square term with the 
largest coefficient can be cancelled by the wino 
mass-square term in $m_{H_u}^2(m_Z)$ with the ratio 
$M_2/M_3 \sim \sqrt{3.1/0.17} \sim 4.3$, that is 
within the range $2 \lesssim M_2/M_3 \lesssim 5$ 
required for $|\tilde{A}_t|/M_{\rm st}>1$.

\section{The Higgs mass and the naturalness}
\label{sec:numerical}
Motivated by the above rough estimations based on 
Eqs.~(\ref{eq:mh}), (\ref{eq:spsatmz}) and (\ref{eq:mhuatmz}), 
we search a parameter space of the MSSM with soft terms, 
especially a region of gaugino mass ratios, where 
the mass of the lightest CP-even Higgs boson $m_h$ resides 
in $124.4-126.8$ GeV and at the same time the degree of 
tuning the $\mu$-parameter is relaxed above 10 \%. 
Because we take $\tan \beta =15$ as declared at the 
end of Appendix~\ref{app:bc}, the effect of the 
other Yukawa couplings than the top one, especially the 
bottom Yukawa coupling, can induce sizable corrections, 
we numerically solve the full set\footnote{Although the 
effects of light generations are negligible, for concreteness, 
we input numerical values of all the Yukawa couplings determined 
by Froggatt-Nielsen mechanism~\cite{Froggatt:1978nt} 
at the GUT scale, shown in Appendix~\ref{app:bc}. 
Also, our calculation covers the situation that the scalar charm quark 
masses are about $10^2$ times higher than scalar top quark masses and the correction 
to $m_{H_u}^2$ running is sizable, as can be seen in Appendix~\ref{app:relRGEs}.} 
of one-loop RGEs including all the Yukawa couplings. 
(The complete set of the MSSM RGEs are found, e.g., 
in Ref.~\cite{Martin:1993zk} at the two-loop level.) 

By utilizing the numerical solutions of RGEs, we evaluate 
the Higgs mass $m_h$ based on the mass matrix with one- 
and two-loop contributions from top and bottom squarks, 
respectively, derived in Ref.~\cite{Carena:1995wu}. 
Here we do not adopt the approximated expression~(\ref{eq:mh}) 
because left- and right-handed top squarks are not degenerate 
in the favored region $2 \lesssim M_2/M_3 \lesssim 5$ 
with the maximal-mixing of top squarks. 
Although the above rough estimation is based on Eq.~(\ref{eq:mh}), 
it turns out to be valid and consequently a heavier Higgs mass can 
be obtained in the full numerical analysis as we will see later. 

We assume a certain mechanism of supersymmetry breaking 
which determines soft supersymmetry breaking parameters 
at the GUT scale, and then the ratios among these parameters 
are fixed with an accuracy. The $\mu$-parameter in the MSSM 
superpotential (6) is in general independent to the mechanism 
of supersymmetry breaking. As a measure of the degree of 
tuning the $\mu$-parameter at the GUT scale, 
we adopt a parameter $\Delta_\mu$ defined by 
\begin{eqnarray}
\Delta_\mu &=& \frac{|\mu|}{2m_Z^2} 
\frac{\partial m_Z^2}{\partial |\mu|}, 
\label{eq:deltamu}
\end{eqnarray}
which represents a sensitivity~\cite{Barbieri:1987fn} 
of the $Z$-boson mass $m_Z$ at the EW scale 
on the $\mu$-parameter at the GUT scale. 
The degree of tuning to obtain $m_Z=91.2$ GeV is then 
estimated as $100 \times |\Delta_\mu^{-1}|$ \%. 
It seems that, with the current experimental status, 
$|\Delta_\mu| > 100$ is inevitable with universal values 
of soft parameters at the GUT scale, which requires 
more severe fine-tuning than the degree of $1$ \%. 

\begin{figure}[t]
\hfill 
\includegraphics[ width=0.45\linewidth]{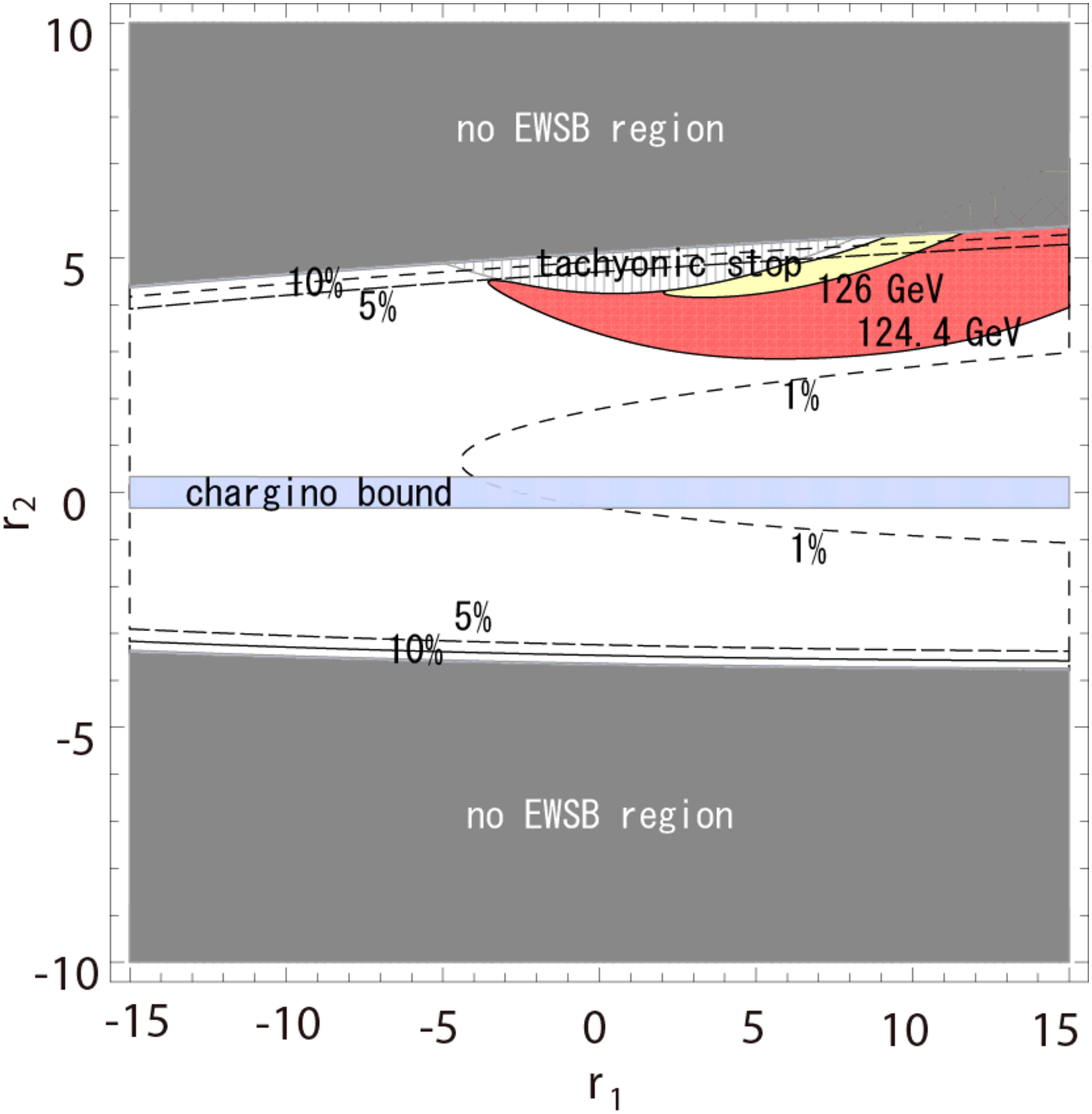}
\hfill 
\includegraphics[width=0.45\linewidth]{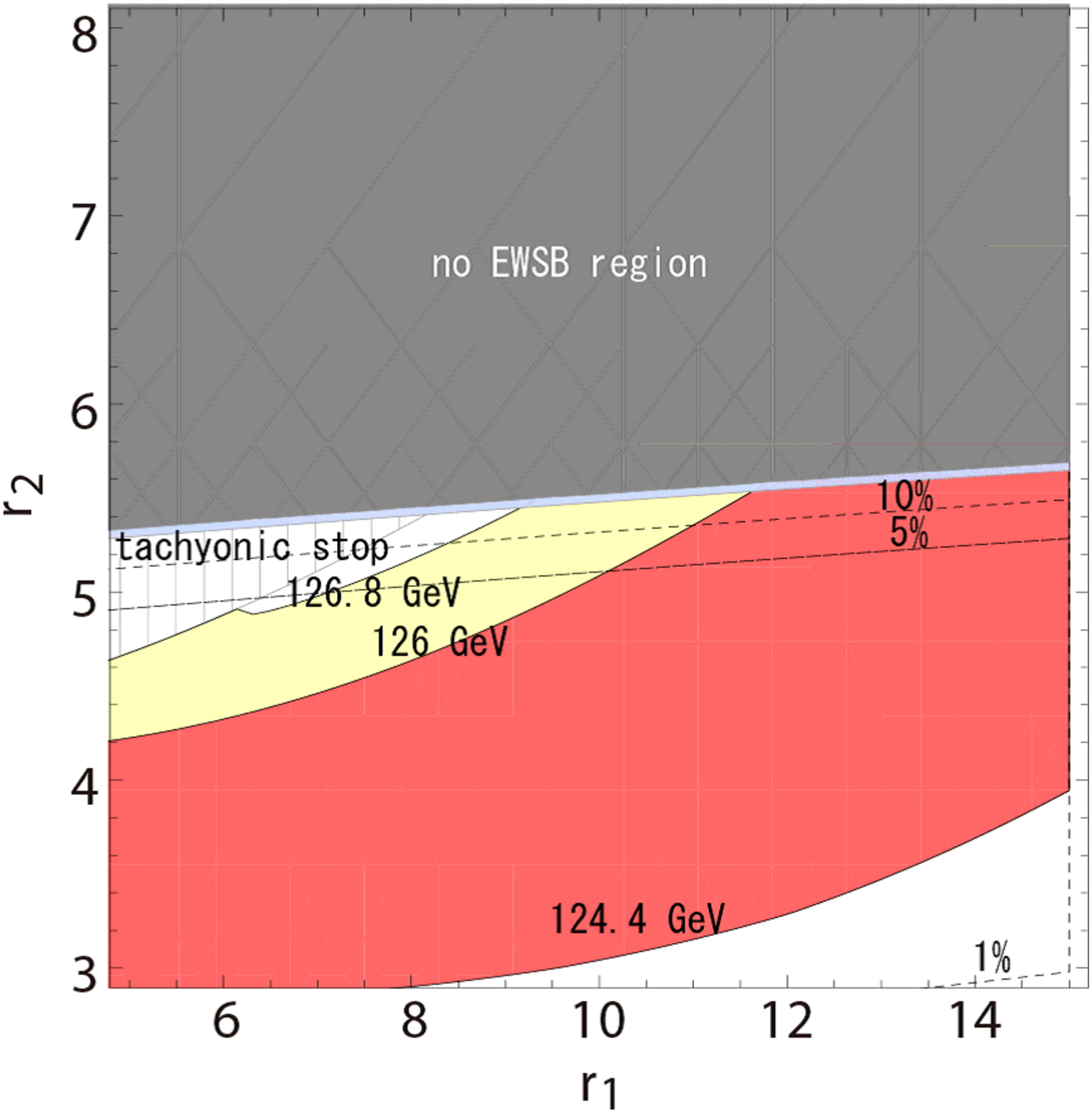}
\hfill 
\caption{Contours of $m_h$ [GeV] and 
$100 \times |\Delta_\mu^{-1}|$ (\%) 
in the parameter space $(r_1,r_2)$.}
\label{fig:mhdmu}
\end{figure}

The results from the direct search at the LHC set 
the lower bounds of the mass of gluino $860$ GeV 
and of the masses of squarks in the first and the 
second generations $1320$ GeV~\cite{:2012rz}. 
We adopt the severest bound although it may be lowered
in models with some nonuniversal gaugino mass ratios~\cite{Caron:2012sf}.
Taking these stringent bounds and the other experimental 
bounds for top squarks, neutralinos and charginos~\cite{Beringer:1900zz} 
into account, we fix numerical values of the other soft 
parameters than wino and bino masses at the GUT scale as 
shown in Appendix~\ref{app:bc}, and vary gaugino mass ratios 
$r_1$, $r_2$ at the GUT scale in the evaluation of the Higgs 
mass $m_h$ and the parameter $|\Delta_\mu|$. 
Fig.~\ref{fig:mhdmu} shows contours of 
$m_h$ [GeV] and $100 \times |\Delta_\mu^{-1}|$ (\%) 
in the parameter space $(r_1,r_2)$. 

From Fig.~\ref{fig:mhdmu}, we find that the mass of the Higgs 
boson resides in $124.4-126.8$ GeV in the region 
$3.0 \lesssim r_2 \lesssim 5.5$ for the wide range of $r_1 \gtrsim -3$. 
Moreover we emphasize that the region is overlapped with 
$5.2 \lesssim r_2 \lesssim 5.5$ where the degree of tuning 
the $\mu$-parameter is relaxed above $10$ \%. 
Therefore we find a parameter region where the Higgs 
boson mass $\sim 125$ GeV indicated by recent LHC 
observations is predicted in a natural MSSM with 
certain nonuniversal gaugino masses at the GUT scale. 
Compared with Ref.~\cite{Abe:2007kf}, here the Higgs boson mass 
above $115$ GeV is achieved by a larger value of $\tan \beta=15$, 
that is one of the reason we employed the full set of MSSM RGEs 
including all the Yukawa couplings. 
Although the overlapped region is not so wide, 
the required accuracy for the ratios of the gaugino 
masses are not so stringent. 
There would exist some supersymmetry breaking and the 
mediation mechanisms that fix the gaugino mass ratios to 
the above numerical values favored by the naturalness. 
For example, concrete fixed values of the ratios are shown 
in Ref.~\cite{Choi:2007ka} for various breaking 
and mediation mechanisms. 

\begin{table}[t]
\begin{center}
\begin{tabular}{|c|c||c|c|} \hline 
sparticle & mass [GeV] & sparticle & mass [GeV] \\ \hline\hline 
$\tilde{u}_1$ &2007&$\tilde{e}_1$ &2119\\ \hline
$\tilde{u}_2$ &2198&$\tilde{e}_2$ &2132\\ \hline
$\tilde{c}_1$ &2002&$\tilde{\mu}_1$ &2104\\ \hline
$\tilde{c}_2$ &2194&$\tilde{\mu}_2$ &2132\\ \hline
$\tilde{t}_1$ &505.9&$\tilde{\tau}_1$ &1492\\ \hline
$\tilde{t}_2$ &1337&$\tilde{\tau}_2$ &1511\\ \hline\hline
$\tilde{d}_1$ &1812 &$\tilde{\chi}_1^0$ &1588\\ \hline
$\tilde{d}_2$ &2200 &$\tilde{\chi}_2^0$ &1558\\ \hline
$\tilde{s}_1$ &1786 &$\tilde{\chi}_3^0$ &176.6\\ \hline
$\tilde{s}_2$ &2196 &$\tilde{\chi}_4^0$ &182.1\\ \hline
$\tilde{b}_1$ &941.8 &$\tilde{\chi}_1^{\pm}$ &179.0\\ \hline
$\tilde{b}_2$ &1317 &$\tilde{\chi}_2^{\pm}$ &1558\\ \hline\hline
\end{tabular}
\caption{A typical superparticle (sparticle) spectrum 
at the EW scale for $(r_1,r_2)=(11,5.4)$. 
The subscripts label the mass eigenstates for 
the up ($\tilde{u}$), down ($\tilde{d}$), charm ($\tilde{c}$), 
strange ($\tilde{s}$), top ($\tilde{t}$), bottom ($\tilde{b}$) squarks, 
the scalar electron ($\tilde{e}$), 
muon ($\tilde{\mu}$), tauon ($\tilde{\tau}$), 
the neutralino ($\tilde\chi^0$) 
and the chargino ($\tilde\chi^{\pm}$).}
\label{tab:spectrum}
\end{center}
\end{table} 

\begin{table}[t]
\begin{center}
\begin{tabular}{|c|c|c|c|} \hline 
$m_h$ [GeV] & $m_H$ [GeV] & $m_A$ [GeV] & $m_{H\pm}$ [GeV] 
\\ \hline
126.0&1415&1415&1417 \\ \hline\hline
$100 \times |\Delta_\mu^{-1}|$ (\%) &$M_1(m_Z)$[GeV] & 
$M_2(m_Z)$[GeV]&$M_3(m_Z)$[GeV] \\ \hline
12.84&1587&1554&1003 \\ \hline
\end{tabular}
\caption{The masses of neutral and charged Higgs bosons 
as well as the gaugino masses at the EW scale for 
$(r_1,r_2)=(11,5.4)$. 
The degree of tuning the $\mu$-parameter, 
$100 \times |\Delta_\mu^{-1}|$ (\%), is also shown.}
\label{tab:higgsmasses}
\end{center}
\end{table} 

\begin{figure}[t]
\centerline{
\includegraphics[width=0.95\linewidth]{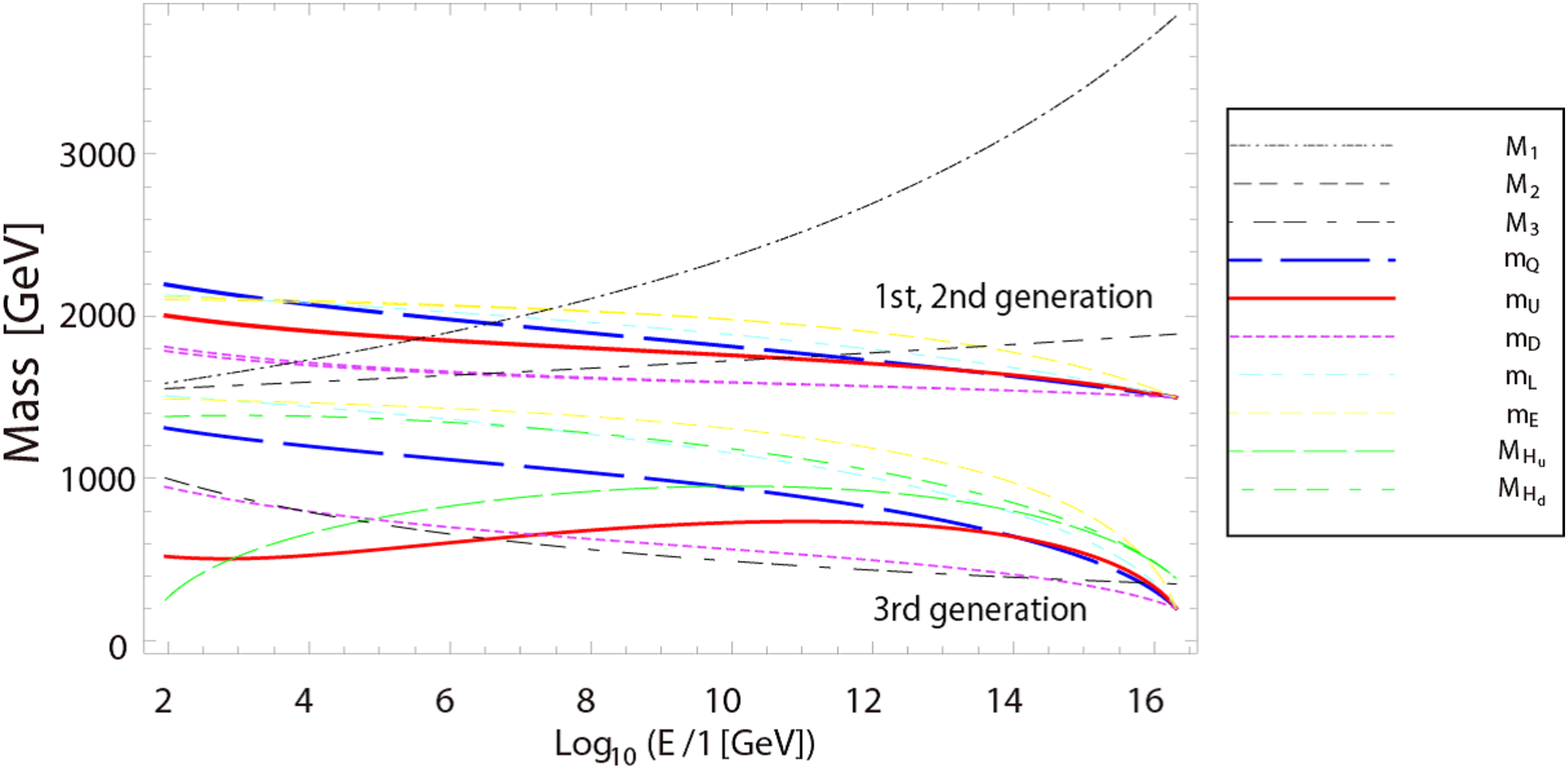}}
\caption{The running of gaugino, soft scalar masses and the parameter 
$M_{H_{u,d}}=(|\mu|^2+m_{H_{u,d}}^2)^\frac{1}{2}$ 
from the GUT to the EW scale for $(r_1,r_2)=(11,5.4)$.}
\label{fig:RG}
\end{figure}

A typical superparticle spectrum at the EW scale and 
the masses of neutral and charged Higgs bosons 
are shown in Tables~\ref{tab:spectrum} and \ref{tab:higgsmasses}, 
by setting the gaugino mass ratios 
$(r_1,r_2)=(11,5.4)$ inside the favored region found above. 
From these tables, we find all the experimental lower bounds 
on the masses of gluinos, neutralinos, charginos, squarks and 
sleptons as well as neutral and charged Higgs bosons can be 
satisfied with $12.84$ \% tuning of the $\mu$-parameter. 
With this parameter choice, Fig.~\ref{fig:RG} 
shows the running of gaugino and soft scalar masses 
from the GUT to the EW scale. We find that gaugino 
masses tend to degenerate at low energies, that is a 
typical signal of this parameter region as discussed in Ref.~\cite{Abe:2007kf}.

Here we comment on charge and 
color breaking minima~\cite{Frere:1983ag}. 
If we take $A_t$ larger than $M_{\rm st}$, 
there is a possibility that such minima appear 
along the direction satisfying 
$|\tilde{q}_3|=|\tilde{u}_3|=|h_u|$ 
in the field space. 
The minima exist unless the following condition, 
\begin{eqnarray}
|A_t|^2 &\leq& 3 \left( 
m_{Q_3}^2 +m_{U_3}^2 +m_{H_u}^2 +|\mu|^2 \right), 
\label{eq:CCB}
\end{eqnarray}
is satisfied. 
It has been noted that the parameters yielding $r_a \simeq {\sqrt 6}$ 
are dangerous with the parameters satisfying $m_{Q_3}=m_{U_3}$ 
due to the inequality~(\ref{eq:CCB})~\cite{Brummer:2012ns}. 
On the other hand, in our analysis, the important region 
${\sqrt 5} \lesssim r_a \lesssim {\sqrt 7}$ shown in Fig.~\ref{fig:atmst} 
appears with the parameters satisfying 
$\left| A_t(m_Z) \right| \sim m_{Q_3}(m_Z)$ 
and $0 < m_{U_3}(m_Z) < m_{Q_3}(m_Z)$, 
and then the inequality~(\ref{eq:CCB}) is guaranteed. 

Finally, we mention about the next to minimal 
supersymmetric standard model (NMSSM). When a singlet chiral 
multiplet is added, the MSSM superpotential can be modified 
to have a positive correction to the Higgs mass square at 
the tree level. There is, however, another negative contribution 
to the Higgs mass square induced by a mixing with the singlet field. 
The latter negative contribution also becomes larger in general 
when the parameters are chosen to make the former positive 
contribution larger. This fact makes the analysis of the NMSSM 
complicated (see for a review, Ref.~\cite{Ellwanger:2009dp}). 
We would not be able to make a definitive statement, 
at the current stage, that the NMSSM is better than the MSSM 
from the viewpoint of both the Higgs mass and the naturalness, 
even though there have been interesting studies concerning 
those issues in the NMSSM~\cite{Cao:2012fz}.

\section{Conclusions and discussions}
\label{sec:conclusion}

By solving the full set of one-loop RGEs numerically 
for all the parameters in the MSSM with soft supersymmetry 
breaking terms, we identified a parameter region where the 
mass of the lightest CP-even Higgs boson resides in 
$124.4-126.8$ GeV indicated by the recent LHC results, 
and at the same time the degree of tuning 
the $\mu$-parameter is relaxed above $10$ \%. 
The region is characterized by certain nonuniversal values 
of gaugino mass parameters at the GUT scale as indicated 
in Ref.~\cite{Abe:2007kf} before the LHC observations. 
We have confirmed that, even after the LHC results, 
the main suggestions in Ref.~\cite{Abe:2007kf} are still 
valid with a larger value of $\tan \beta$ based on more 
accurate analyses than those in Ref.~\cite{Abe:2007kf}. 

We also derived a superparticle spectrum for a typical 
parameter choice inside the identified region. 
We find all the experimental lower bounds on the masses of 
gluinos, neutralinos, charginos, squarks and sleptons 
as well as neutral and charged Higgs bosons can be satisfied. 
One of the outstanding features of the spectrum is the 
degenerate gaugino masses at low energies as a consequence 
of a particular choice of gaugino mass ratios at the GUT scale 
favored by the naturalness, as mentioned in Ref.~\cite{Abe:2007kf}. 
It would be interesting to study cosmological features 
in detail in this parameter region of the MSSM\footnote{ 
A possibility of the neutralino dark matter was studied in 
Ref.~\cite{Abe:2007je} based on Ref.~\cite{Abe:2007kf} 
before the LHC results.}$^,$\footnote{
Candidates for a dark matter in this region are the gravitino, 
the right-handed sneutrino, and the Higgsino-like neutralino. 
In the third case, the Higgsino-like neutralino dark matter 
would be possible due to an enhancement from a non-thermal decay 
(gravitino \cite{Kohri:2005ru}, Axino \cite{Choi:2008zq} 
or moduli decay \cite{Moroi:1999zb}), 
although the thermal abundance of Higgsino-like neutralino is 
sub-dominant to explain the current dark matter abundance.}. 

One of the important guiding principles for the physics 
beyond the MSSM is provided by the argument of its naturalness. 
The EW scale is unstable under a tiny numerical deviation 
of the $\mu$-parameter when a severe tuning is required 
to realize the observed mass of $W$ and $Z$ bosons. 
Therefore the underlying theory free from such a fine-tuning 
is desirable, which becomes a strong guiding principle when 
we study particle physics models at a more fundamental level. 
We adopted Eq.~(\ref{eq:deltamu}) as the measure of the degree 
of tuning. The measure can be extended to include all the other
soft supersymmetry breaking parameters like, e.g., in Ref.~\cite{Cassel:2010px}. 
However, we have assumed that the ratio between the breaking 
parameters are fixed with an enough accuracy by a concrete 
supersymmetry breaking and the mediation mechanism. 
Even in this situation, the supersymmetric parameter $\mu$ 
would be independent to the other parameters, and the issues 
of the fine-tuning should be concerned about the $\mu$-parameter.

There would exist some supersymmetry breaking and the mediation 
mechanisms that fix the gaugino mass ratios~\cite{Choi:2007ka} 
to the favored numerical values by the naturalness. 
One of such candidates is a mirage mediation model~\cite{Endo:2005uy}, 
where the gaugino mass ratio at the GUT scale is determined 
by the ratio of contributions to gaugino masses from the modulus 
and the anomaly mediated supersymmetry breaking~\cite{Choi:2004sx}. 
Issues of fine-tuning in this model are studied in Ref.~\cite{Choi:2005hd}. 
Another and more general origin of nonuniversal gaugino masses 
is moduli-mixing gauge kinetic functions, which appear even at 
the tree level in a certain effective supergravity action. 
For example, in some superstring models, such moduli-mixings 
appear from nontrivial D-brane configurations where the gaugino 
mass ratios are determined by, e.g., numbers of windings, 
intersections and magnetic fluxes of 
D-branes~\cite{Lust:2004cx} (see, for a reivew, Ref.~\cite{Blumenhagen:2006ci}). 
Even without mentioning superstrings, in a minimal extension of 
the MSSM with a single extra dimension, namely, in five-dimensional 
supergravity models, the situation is similar and the gaugino 
mass ratios can be determined by data of the very special manifold 
governing the structure of ${\cal N}=2$ vector 
multiplets~\cite{Ceresole:2000jd}. 
In all the cases with moduli-mixing gauge kinetic functions, 
the mechanism of moduli stabilization is important to determine 
the gaugino mass ratios~\cite{Abe:2005rx}. 

The LHC is now exploring the parameter space of the MSSM 
and the other supersymmetric models. 
It would be possible that the recent and near-future 
observations guide us in a direction toward a more fundamental 
theory of the nature through the results obtained here.

\subsection*{Acknowledgement}
The authors would like to thank T.~Higaki for 
stimulating discussions. The work of H.~A. was 
supported by the Waseda University Grant for 
Special Research Projects No.2012B-151.

\appendix

\section{Boundary conditions at the GUT scale}
\label{app:bc}

In this appendix we show numerical values of parameters 
in the MSSM with soft supersymmetry breaking terms 
selected as a boundary condition at the GUT scale, 
$M_{\rm GUT}=2 \times 10^{16}$ GeV, 
for our numerical analysis of the full set of 
MSSM one-loop RGEs. 

For the chiral multiplets 
$Q_i$, $U_i$, $D_i$, $L_i$ and $E_i$ 
carrying the $i$th generation of 
the left-handed quark doublet, 
the right-handed up quark,  
the right-handed down quark,  
the left-handed lepton doublet and 
the right-handed electron, 
and those $H_u$ and $H_d$ containing the up- 
and the down-type Higgs doublet, respectively, 
the MSSM superpotential is given by 
\begin{eqnarray}
W_{\rm MSSM} &=& \mu H_u H_d 
+y^u_{ij}H_u Q_i U_j 
+y^d_{ij}H_d Q_i D_j 
+y^e_{ij}H_d L_i E_j, 
\label{eq:wmssm}
\end{eqnarray}
where $y^{u,d,e}_{ij}$ are Yukawa coupling constants, 
and $\mu$ is the Higgsino mass parameter referred to 
as $\mu$-parameter. The summations over the generation 
indices $i,j=1,2,3$ are implicit. 
The soft supersymmetry breaking terms are defined by 
\begin{eqnarray}
{\cal L}_{\rm soft} &=& 
-\frac{1}{2} \left( \sum_{a=1}^3 M_a 
{\rm tr}\,\lambda^a \lambda^a +{\rm h.c.} \right) 
\nonumber \\ &&
-\sum_\Phi (m_\Phi^2)_{ij} 
\tilde\varphi_i^\dagger \tilde\varphi_j 
-m_{H_u}^2 |h_u|^2 -m_{H_d}^2 |h_d|^2 
\nonumber \\ && - \left( 
 a^u_{ij}h_u \tilde{q}_i \tilde{u}_j 
+a^d_{ij}h_d \tilde{q}_i \tilde{d}_j 
+a^e_{ij}h_d \tilde{l}_i \tilde{e}_j 
+B \mu h_u h_d +{\rm h.c.} \right), 
\label{eq:softterms}
\end{eqnarray}
where 
$\tilde\varphi = \tilde{q}, \tilde{u}, \tilde{d}, 
\tilde{l}, \tilde{e}$ and $h_u, h_d$ 
represent the scalar components of the chiral multiplets 
$\Phi = Q, U, D, L, E$ and $H_u, H_d$, respectively, 
and $\lambda^3$, $\lambda^2$ and $\lambda^1$ are gaugino 
fields in the vector multiplets for 
$SU(3)_C$, $SU(2)_L$ and $U(1)_Y$ 
gauge groups of the MSSM, respectively. 
The gaugino masses $M_a$, 
the soft scalar masses $(m_\Phi^2)_{ij}$ and 
the scalar trilinear couplings $a^{u,d,e}_{ij}$ 
as well as the parameter $B \mu$ in Eq.~(\ref{eq:softterms})
are called soft supersymmetry breaking parameters. 
Note that the MSSM singlet chiral multiplet carrying 
a right-handed neutrino could be introduced 
without affecting the basic results of this paper, 
which is omitted just for simplicity. 

As for supersymmetric parameters, 
three gauge coupling constants $g_3$, $g_2$ and $g_1$ 
for $SU(3)_C$, $SU(2)_L$ and $U(1)_Y$ gauge groups in 
the MSSM, respectively, are given at the GUT scale as 
\begin{eqnarray}
g_3 &=& 0.720, \qquad 
g_2 \ = \ 0.719, \qquad 
g_1 \ = \ 0.719, 
\nonumber
\end{eqnarray}
determined by the observed values at low energies. 
The Yukawa coupling matrices at the GUT scale are 
chosen for generation indices $i,j=1,2,3$ as 
\begin{eqnarray}
y^u_{ij} &=& 
\left( \begin{array}{ccc}
0.963 \times \epsilon^{5} 
& 0.457\times\epsilon^{3.5}  
& 0.397 \times \epsilon^{2.5} \\
0.546 \times \epsilon^{4}  
& 0.481\times \epsilon^{2.5} 
& 0.670 \times \epsilon^{1.5} \\ 
0.153 \times\epsilon^{2.5} 
& 0.334\times \epsilon^{1} 
& 0.595 \times \epsilon^{0} 
\end{array} \right), 
\nonumber \\
y^d_{ij} &=& 
\left( \begin{array}{ccc}
0.294 \times\epsilon^{4} 
& 0.496\times \epsilon^{4.5} 
& 0.468 \times\epsilon^{3.5}  \\ 
0.156 \times \epsilon^{3}  
& 0.172\times \epsilon^{3.5}  
& 0.304 \times \epsilon^{2.5}  \\ 
0.527 \times\epsilon^{1.5} 
& 0.775\times\epsilon^{2} 
& 0.456 \times \epsilon^{1} 
\end{array} \right), 
\nonumber \\
y^e_{ij} &=& 
\left( \begin{array}{ccc}
0.573 \times \epsilon^{6} 
& 0.404\times \epsilon^{4} 
& 0.946 \times\epsilon^{4}  \\ 
0.404 \times \epsilon^{5} 
& 1.02\times\epsilon^{3} 
& 0.274 \times \epsilon^{3}  \\ 
0.686 \times \epsilon^{3.5} 
& 0.690\times \epsilon^{1.5} 
& 0.718 \times \epsilon^{1.5}   
\end{array} \right), 
\nonumber
\end{eqnarray}
which can be realized by the 
Froggatt-Nielsen mechanism~\cite{Froggatt:1978nt} 
or quasi-localized matter fields 
in five-dimensional spacetime~\cite{ArkaniHamed:1999dc, Kaplan:2000av} 
yielding observed masses and mixings of quarks 
and charged leptons at the EW scale. 
Here $\epsilon=0.225$ represents 
the magnitude of mixing by Cabibbo angle. 
Note that these numerical values of the elements 
involving light generations are just for 
concreteness of the numerical evaluation, 
and these concrete values for the light generations 
are not essential for our results. 
The basic results are valid for the other ansatz 
of Yukawa matrices that generate observed quark 
and lepton masses and mixings as long as 
it coincides with a value of $\tan \beta \gtrsim 10$ 
as will be selected in Eq.~(\ref{eq:tanbeta}). 

The parameters in soft supersymmetry breaking terms 
are selected as follows. The gluino mass is chosen as 
\begin{eqnarray}
M_3 &=& 350 \textrm{\ GeV}, 
\nonumber
\end{eqnarray}
which is almost the lowest value satisfying the 
experimental lower bound at the EW scale. 
Note that the gaugino mass ratios $r_1$ and $r_2$, 
namely the bino mass $M_1$ and the wino masses $M_2$, 
are varied at the GUT scale with the above fixed 
gluino mass $M_3$ in our numerical analysis. 
The soft scalar masses and the scalar 
trilinear couplings at the GUT scale are fixed as 
\begin{eqnarray}
\sqrt{(m_\Phi^2)_{ij}} &=& \left\{ 
\begin{array}{rcl}
1500 &\textrm{GeV}& (i=j=1,2) \\*[5pt] 
200 &\textrm{GeV}& (i=j=3) \\*[5pt] 
0 &\textrm{GeV}& (i \ne j) \nonumber
\end{array} \right., \qquad 
\Phi=Q,U,D,L,E, \nonumber \\ \nonumber \\ 
m_{H_u} &=& m_{H_d}~=~200~\textrm{GeV},
\nonumber
\end{eqnarray}
and 
\begin{eqnarray}
a^{u,d,e}_{ij} &=& y^{u,d,e}_{ij} A^{u,d,e}_{ij}, \qquad 
A^{u,d,e}_{ij} \ = \ -400 \textrm{\ GeV}, \qquad 
i,j=1,2,3. 
\nonumber
\end{eqnarray}
Although we adopt vanishing off-diagonal elements 
of the soft scalar mass matrices for simplicity, 
these essentially do not affect the main results 
of this paper. 

We choose the following ratio: 
\begin{eqnarray}
\tan \beta &\equiv& v_u/v_d =15, 
\label{eq:tanbeta}
\end{eqnarray}
where $v_u=v \sin \beta$ and $v_d=v \cos \beta$ are 
the VEVs of up- and down-type Higgs fields, $h_u$ and $h_d$, 
respectively, and $v = 174$ GeV. 
The smaller value of $\tan \beta$ makes the Higgs mass 
tend to be below 120 GeV in the MSSM. 
The larger value makes the hierarchy between $v_u$ and 
$v_d$ severer, that is unfavorable form the viewpoint 
of the naturalness. 
The $\mu$-parameter in the MSSM superpotential~(\ref{eq:wmssm}) 
as well as the $B \mu$-parameter in the soft supersymmetry 
breaking terms~(\ref{eq:softterms}) is determined at the GUT scale 
in such a way that the correct $Z$ boson mass is realized 
at the EW scale with the above set of the other parameters. 
Note that the numerical values of these parameters 
changes as $r_1$ and $r_2$ vary.

\section{RGEs relevant to the Higgs mass}
\label{app:relRGEs}

In this appendix, we show the one-loop RGEs relevant to our 
discussion for the mass of the lightest CP-even Higgs boson: 
\begin{eqnarray}
\frac{d m_Q^2}{dt} &=& 
-\frac{1}{4\pi^2} \left( 
\frac{8}{3}g_3^2|M_3|^2 
+\frac{3}{2}g_2^2|M_2|^2 
+\frac{1}{30}g_1^2|M_1|^2 
-\frac{1}{20}g_1^2 S \right) \hat{\bm{1}} 
\nonumber \\ &&
+ \frac{1}{8\pi^2} \left(
\frac{1}{2} y^u (y^u)^\dagger m_Q^2 
+\frac{1}{2} m_Q^2 y^u (y^u)^\dagger 
+y^u m_U^2 (y^u)^\dagger 
+(m_{H_u}^2) y^u (y^u)^\dagger 
+A^u (A^u)^\dagger \right) 
\nonumber \\ &&
+ \frac{1}{8\pi^2} \left( 
\frac{1}{2} y^d (y^d)^\dagger m_Q^2 
+\frac{1}{2} m_Q^2 y^d (y^d)^\dagger 
+y^d m_D^2 (y^d)^\dagger 
+(m_{H_d}^2) y^d (y^d)^\dagger 
+A^d (A^d)^\dagger \right), 
\nonumber \\
\frac{d m_U^2}{dt} &=& 
-\frac{1}{4\pi^2} \left( 
\frac{8}{3}g_3^2|M_3|^2 
+\frac{8}{15}g_1^2|M_1|^2 
+\frac{1}{5}g_1^2 S \right) \hat{\bm{1}} 
\nonumber \\ &&
+ \frac{1}{4\pi^2} \left(
\frac{1}{2} (y^u)^\dagger y^u m_U^2 
+\frac{1}{2} m_U^2 (y^u)^\dagger y^u 
+(y^u)^\dagger m_Q^2 y^u 
+(m_{H_u}^2) (y^u)^\dagger y^u 
+(A^u)^\dagger A^u \right), 
\nonumber \\
\frac{d m_{H_u}^2}{dt} &=& 
-\frac{1}{4\pi^2} \left( 
\frac{3}{2}g_2^2|M_2|^2 
+\frac{3}{10}g_1^2|M_1|^2 
-\frac{3}{20}g_1^2 S\right) 
\nonumber \\ &&
+ \frac{3}{8\pi^2} {\rm tr} 
\left\{ 
y^u m_Q^2 (y^u)^\dagger 
+y^u m_U^2 (y^u)^\dagger 
+m_{H_u}^2 y^u (y^u)^\dagger 
+A^u (A^u)^\dagger \right\}, 
\nonumber \\
S &=& m_{H_u}^2 -m_{H_d}^2 
+{\rm tr} \left( m_Q^2 -m_L^2 -2 m_U^2 
+m_D^2 +m_E^2 \right), 
\nonumber
\end{eqnarray} 
where 
$t = {\rm ln} \left( E/M_{\rm GUT} \right)$ 
is a logarithmic energy scale measured by 
$M_{\rm GUT}=2 \times 10^{16}$ GeV, and 
$\hat{\bm{1}}$ is a $3 \times 3$ unit matrix. 
The complete set of the MSSM RGEs are found, 
e.g., in Ref.~\cite{Martin:1993zk} 
at the two-loop level.


\begin{thebibliography}{99}

\bibitem{Martin:1997ns}
  S.~P.~Martin,
  In *Kane, G.L. (ed.): Perspectives on supersymmetry II* 1-153
  [hep-ph/9709356].

\bibitem{:2012rz}
  G.~Aad {\it et al.}  [ATLAS Collaboration],
  arXiv:1208.0949 [hep-ex].

\bibitem{:2012gk}
  G.~Aad {\it et al.}  [ATLAS Collaboration],
  Phys.\ Lett.\ B
  [arXiv:1207.7214 [hep-ex]];
  S.~Chatrchyan {\it et al.}  [CMS Collaboration],
  Phys.\ Lett.\ B
  [arXiv:1207.7235 [hep-ex]].


\bibitem{Okada:1990vk}
  Y.~Okada, M.~Yamaguchi and T.~Yanagida,
  Prog.\ Theor.\ Phys.\  {\bf 85} (1991) 1.

\bibitem{Abe:2007kf}
  H.~Abe, T.~Kobayashi and Y.~Omura,
  Phys.\ Rev.\ D {\bf 76} (2007) 015002
  [hep-ph/0703044 [hep-ph]].
  
\bibitem{Antusch:2012gv}
  S.~Antusch, L.~Calibbi, V.~Maurer, M.~Monaco and M.~Spinrath,
  arXiv:1207.7236 [hep-ph].

\bibitem{Carena:1995wu}
  M.~S.~Carena, M.~Quiros and C.~E.~M.~Wagner,
  Nucl.\ Phys.\ B {\bf 461} (1996) 407
  [hep-ph/9508343].

\bibitem{Froggatt:1978nt}
  C.~D.~Froggatt and H.~B.~Nielsen,
  Nucl.\ Phys.\ B {\bf 147} (1979) 277.

\bibitem{Martin:1993zk}
  S.~P.~Martin and M.~T.~Vaughn,
  Phys.\ Rev.\ D {\bf 50} (1994) 2282
   [Erratum-ibid.\ D {\bf 78} (2008) 039903]
  [hep-ph/9311340].

\bibitem{Barbieri:1987fn}
  R.~Barbieri and G.~F.~Giudice,
  Nucl.\ Phys.\ B {\bf 306} (1988) 63.

\bibitem{Caron:2012sf}
  S.~Caron, J.~Laamanen, I.~Niessen and A.~Strubig,
  JHEP {\bf 1206} (2012) 008
  [arXiv:1202.5288 [hep-ph]].

\bibitem{Beringer:1900zz}
  J.~Beringer {\it et al.}  [Particle Data Group Collaboration],
  Phys.\ Rev.\ D {\bf 86} (2012) 010001;
  G.~Aad {\it et al.}  [ATLAS Collaboration],
  arXiv:1208.1447 [hep-ex].

\bibitem{Choi:2007ka}
  K.~Choi and H.~P.~Nilles,
  JHEP {\bf 0704} (2007) 006
  [hep-ph/0702146 [hep-ph]].

\bibitem{Frere:1983ag}
  J.~M.~Frere, D.~R.~T.~Jones and S.~Raby,
  Nucl.\ Phys.\ B {\bf 222} (1983) 11; 
  J.~A.~Casas, A.~Lleyda and C.~Munoz,
  Nucl.\ Phys.\ B {\bf 471} (1996) 3 
  [hep-ph/9507294].
  
\bibitem{Brummer:2012ns}
  F.~Brummer, S.~Kraml and S.~Kulkarni,
  JHEP {\bf 1208} (2012) 089
  [arXiv:1204.5977 [hep-ph]].

\bibitem{Ellwanger:2009dp}
  U.~Ellwanger, C.~Hugonie and A.~M.~Teixeira,
  Phys.\ Rept.\  {\bf 496} (2010) 1
  [arXiv:0910.1785 [hep-ph]].

\bibitem{Cao:2012fz}
  J.~-J.~Cao, Z.~-X.~Heng, J.~M.~Yang, Y.~-M.~Zhang and J.~-Y.~Zhu,
  JHEP {\bf 1203} (2012) 086
  [arXiv:1202.5821 [hep-ph]];
%
  M.~Asano and T.~Higaki,
  arXiv:1204.0508 [hep-ph]; 
%
  T.~Kobayashi, H.~Makino, K.~-i.~Okumura, T.~Shimomura and T.~Takahashi,
  arXiv:1204.3561 [hep-ph].

  \bibitem{Abe:2007je}
  H.~Abe, Y.~G.~Kim, T.~Kobayashi and Y.~Shimizu,
  JHEP {\bf 0709} (2007) 107
  [arXiv:0706.4349 [hep-ph]].

  \bibitem{Kohri:2005ru}
  K.~Kohri, M.~Yamaguchi and J.~'i.~Yokoyama,
   Phys.\ Rev.\ D {\bf 72} (2005) 083510
   [hep-ph/0502211].

\bibitem{Choi:2008zq}
  K.~-Y.~Choi, J.~E.~Kim, H.~M.~Lee and O.~Seto,
  Phys.\ Rev.\ D {\bf 77} (2008) 123501
  [arXiv:0801.0491 [hep-ph]].

\bibitem{Moroi:1999zb}
  T.~Moroi and L.~Randall,
  Nucl.\ Phys.\ B {\bf 570} (2000) 455
  [hep-ph/9906527].
 

\bibitem{Cassel:2010px}
  S.~Cassel, D.~M.~Ghilencea and G.~G.~Ross,
  Nucl.\ Phys.\ B {\bf 835} (2010) 110
  [arXiv:1001.3884 [hep-ph]];
  D.~M.~Ghilencea, H.~M.~Lee and M.~Park,
  JHEP {\bf 1207} (2012) 046
  [arXiv:1203.0569 [hep-ph]].



\bibitem{Endo:2005uy}
  M.~Endo, M.~Yamaguchi and K.~Yoshioka,
  Phys.\ Rev.\ D {\bf 72} (2005) 015004
  [hep-ph/0504036]; 
  K.~Choi, K.~S.~Jeong and K.~-i.~Okumura,
  JHEP {\bf 0509} (2005) 039
  [hep-ph/0504037].

\bibitem{Choi:2004sx}
  K.~Choi, A.~Falkowski, H.~P.~Nilles, M.~Olechowski and S.~Pokorski,
  JHEP {\bf 0411} (2004) 076
  [hep-th/0411066]; 
  K.~Choi, A.~Falkowski, H.~P.~Nilles and M.~Olechowski,
  Nucl.\ Phys.\ B {\bf 718} (2005) 113
  [hep-th/0503216].

\bibitem{Choi:2005hd}
  K.~Choi, K.~S.~Jeong, T.~Kobayashi and K.~-i.~Okumura,
  Phys.\ Lett.\ B {\bf 633} (2006) 355
  [hep-ph/0508029]; 
  R.~Kitano and Y.~Nomura,
  Phys.\ Lett.\ B {\bf 631} (2005) 58
  [hep-ph/0509039].]

\bibitem{Lust:2004cx}
  D.~Lust, P.~Mayr, R.~Richter and S.~Stieberger,
  Nucl.\ Phys.\ B {\bf 696} (2004) 205
  [hep-th/0404134].

\bibitem{Blumenhagen:2006ci}
  R.~Blumenhagen, B.~Kors, D.~Lust and S.~Stieberger,
  Phys.\ Rept.\  {\bf 445} (2007) 1
  [hep-th/0610327].

\bibitem{Ceresole:2000jd}
  A.~Ceresole and G.~Dall'Agata,
  Nucl.\ Phys.\ B {\bf 585} (2000) 143
  [hep-th/0004111].

\bibitem{Abe:2005rx}
  H.~Abe, T.~Higaki and T.~Kobayashi,
  Phys.\ Rev.\ D {\bf 73} (2006) 046005
  [hep-th/0511160]; 
  Nucl.\ Phys.\ B {\bf 742} (2006) 187
  [hep-th/0512232].

\bibitem{ArkaniHamed:1999dc}
  N.~Arkani-Hamed and M.~Schmaltz,
  Phys.\ Rev.\ D {\bf 61} (2000) 033005
  [hep-ph/9903417].

\bibitem{Kaplan:2000av}
  D.~E.~Kaplan and T.~M.~P.~Tait,
  JHEP {\bf 0006} (2000) 020
  [hep-ph/0004200].

\end{thebibliography}
\end{document}